\documentclass[12pt]{iopart}
\usepackage{float}
\usepackage{subfig}
\usepackage[utf8]{inputenc}
\usepackage{graphicx}
\usepackage{dcolumn}
\usepackage{bm}
\usepackage{hyperref}

\begin{document}

\title{Orbital angular momentum tuning using a phase only parallel aligned LCoS display}

\author{M Vergara$^{1,2}$ and C Iemmi$^{1,2}$}

\address{$^{1}$ Facultad de Ciencias Exactas y Naturales, Departamento de Física, Universidad de Buenos Aires, Buenos Aires, Argentina.}
\address{$^{2}$ Consejo Nacional de Investigaciones Científicas y Técnicas, Buenos Aires, Argentina.}

\ead{marto@df.uba.ar}

\begin{abstract}
    We propose two schemes for achieving continuous and complete orbital angular momentum tuning using a simple device based on a phase only parallel aligned liquid crystal on silicon display, where the tuning is managed by changing the input beam state of polarization. We use an optical tweezers toolbox to simulate the optical force field of the tuned beams applied to dielectric spherical particles in order to prove the applicability of the method.
\end{abstract}

\maketitle

Since the discovery in 1992 \cite{allen92} that light with an azimuthally varying phase like $\exp(il\theta)$, with integer $l$, carried orbital angular momentum (OAM) of $l \hbar$ per photon [arbitrarily higher than spin angular momentum (SAM)] there has been a huge effort oriented to generate, characterize and apply these kind of light, in different research fields. Orbital contribution to light's angular momentum has given a new degree of freedom with a great variety of applications, from quantum information \cite{schmiegelow12} and microscopy \cite{furhapter05} to astrophysics \cite{thide11}. One of the fundamental properties of light's OAM is that it can be transferred to dielectric particles in the form of a mechanical torque \cite{novotny12}, providing the possibility to rotate optically trapped particles, which gives important advances in micromanipulation \cite{bowman11}, and in the design and controlling of micromachines \cite{asavei09}.

Several approaches have been proposed for experimental generation of beams carrying OAM \cite{yao11}. However, continuous controlled OAM tuning is still not a simple task and has attracted attention in recent years. The main challenge is to achieve continuous arbitrary modulation of the OAM of a beam leaving unaltered the intensity distribution, and keeping in every case the stability of the beam during propagation. Techniques like superposition of opposite handedness vortexes \cite{schmitz06} and creation of beams with fractional OAM \cite{vega07} are unsuccessful in this sense, since the OAM change goes hand in hand with a change in the intensity distribution and a loss of symmetry, making these methods inconvenient for example for optical trapping. In a more recent approach M. Gecevičius \textit{et al.} use a super-structured half-wave plate polarization converter to control the OAM by controlling the spin angular momentum of the input light \cite{gecevicius14}. Also there have been used a device based on a spatial light modulator (SLM) to create 2 beams with orthogonal polarization and inverse OAM, achieving a complete OAM tuning by attenuating one of them and inverting the phase addressed onto the SLM \cite{pan16}.

In this letter we propose two simple schemes for achieving continuous and arbitrary OAM tuning based on a commercially available parallel aligned liquid crystal on silicon (PA-LCoS) display, where the OAM tuning is achieved by varying the input light's polarization angle or ellipticity, respectively. We numerically simulate the optical forces involved and show how this can be applied to optical trapping and micromanipulation.

\begin{figure}[H]
    \centering
    \includegraphics[width=0.5\columnwidth]{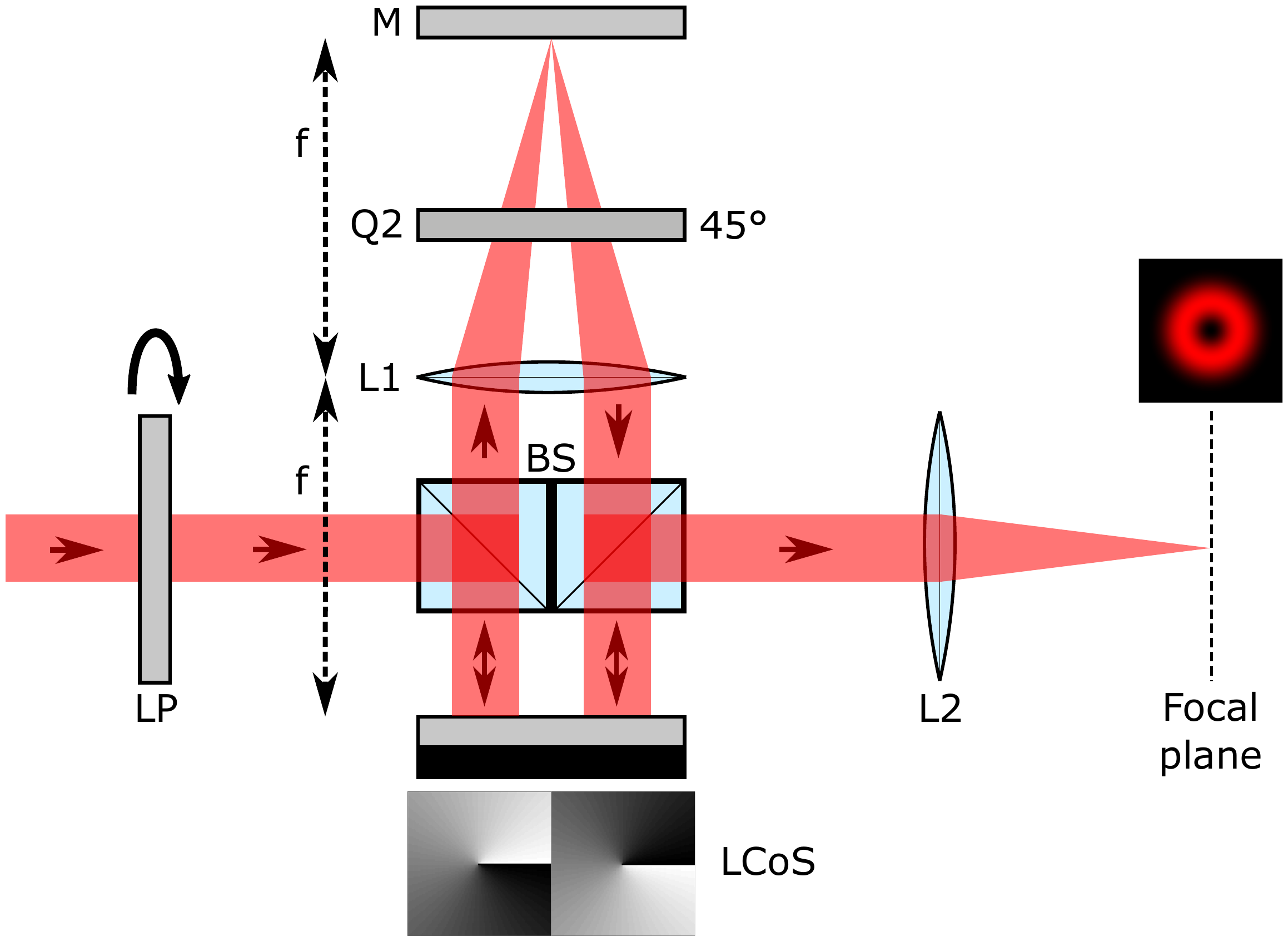}
    \caption{Experimental setup proposed for arbitrary controlled OAM tuning. LP is a linear polarizer, Q2 is a quarter wave plate, Ls are convergent lenses, M is a mirror and BS is a pair of beam-splitters. Horizontal and vertical polarization components of the input beam acquire helical phase profiles of opposite handedness, which are programmed onto a phase only LCoS.}
    \label{fig:dispositivo1}
\end{figure}

The first proposed device is sketched in Fig. \ref{fig:dispositivo1}. An input beam, polarized according to the orientation angle of a linear polarizer LP, is reflected by a first beam splitter BS and impinges onto one half of the LCoS, which introduces a phase modulation on the horizontal component of the field (let us suppose that the director of the LC molecules is horizontally oriented). This half is programmed with a phase modulation of $\psi = m\theta$, where $\theta$ is the azimuthal coordinate and $m$ is any integer number. Then, the reflected beam is imaged by means of a lens L1 of focal length \textbf{f}, and a mirror M (both conforming a 4f system) onto the second half of the LCoS. The quarter wave plate Q2, oriented at $45^\circ$, rotates the polarization angle $90^\circ$ due to the double passage. This way, the initially vertical polarization becomes horizontal and is modulated by the phase $-\psi = -m\theta$ encoded on this second half. The output field is redirected by a second BS and focused by means of a lens L2. The Jones matrix describing the effect of the device is then
\begin{eqnarray}
    J(\theta) &= D_{-m}(\theta)\cdot Q(-45^\circ)\cdot M\cdot Q(45^\circ)\cdot D_{m}(\theta),
\end{eqnarray}
where $D_{m}(\theta)$, $Q(\phi)$ and $M$ are the Jones matrices of the LCoS display programmed with a phase $m\theta$, a quarter wave plate oriented at an angle $\phi$ \cite{goldstein03}, and a plane mirror, respectively. Since these matrix expressions are given by
\begin{eqnarray}
    &D_{m}(\theta) =
    \left( \begin{array}{cc}
    \exp(im\theta) & 0 \\
    0 & 1 \end{array} \right), \\
    &Q(\phi) = 
    \left( \begin{array}{cc}
    \frac{1}{\sqrt{2}}+\frac{i}{\sqrt{2}}\cos(2\phi) & \frac{i}{\sqrt{2}}\sin(2\phi) \\
    \frac{i}{\sqrt{2}}\sin(2\phi) & \frac{1}{\sqrt{2}}-\frac{i}{\sqrt{2}}\cos(2\phi) \end{array} \right),\\
    &M = 
    \left( \begin{array}{cc}
    1 & 0 \\
    0 & -1 \end{array} \right),
\end{eqnarray}
the resulting matrix of the device is
\begin{eqnarray}
    J(\theta) =
    \left( \begin{array}{cc}
    0 & -i\exp(-im\theta) \\
    i\exp(im\theta) & 0 \end{array} \right).
\end{eqnarray}

This way, the horizontal component of the input beam is turned into vertical and emerges with a phase factor $\exp(im\theta)$, while vertical component of the input field is turned into horizontal, with a phase factor $\exp(-im\theta)$. The use of this LCoS display allows changing the addressed phase at will, without requiring to use different elements to reach different OAM values.

When the input beam is linearly polarized, the polarization angle gives the ratio between the horizontal and vertical components of the electric field, and in consequence the ratio between the intensity of the output beam carrying OAM with $l=m$ and that carrying OAM with $l=-m$. Thereby, in the proposed scheme, it is possible to achieve a continuous variation of the total OAM between $-m \leq l \leq m$ by rotating the orientation of the linear polarizer LP.

Since the OAM topological charges of both polarization components are integer and opposite, they show the same intensity distribution, stable through propagation. And since orthogonal polarization components do not interfere, this intensity distribution is unaffected when tuning the OAM.

\begin{figure}[H]
    \centering
    \includegraphics[width=0.5\columnwidth]{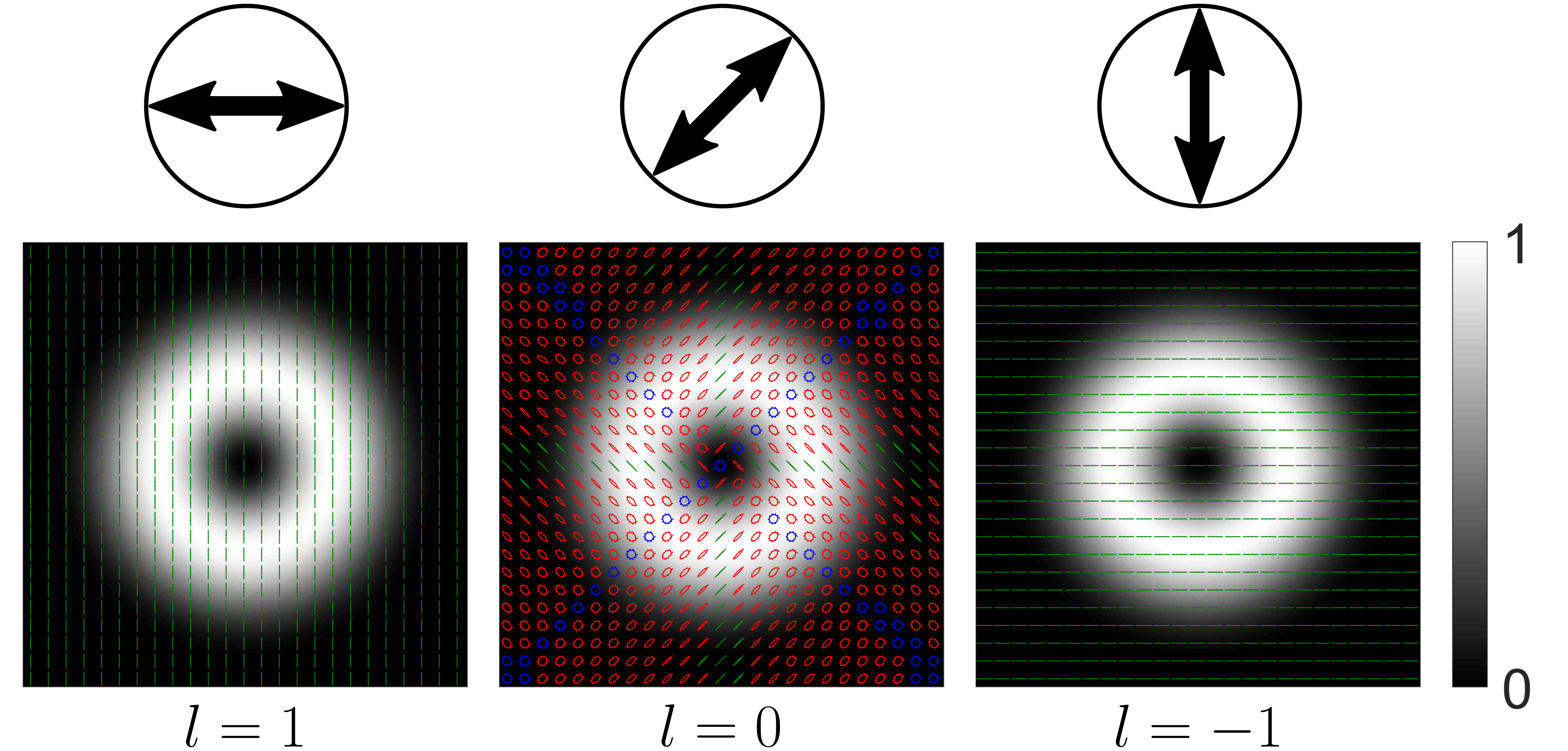}
    \caption{OAM tuning with a linear polarization basis. Top: SoP of the incident light. Middle: Intensity and polarization structure of the OAM tuned beam. Bottom: OAM topological charge. Intensity distribution remains, while polarization changes its structure. In Visualization 1 we show the complete evolution of the output SoP.}
    \label{fig:polarizacion}
\end{figure}

In Fig. \ref{fig:polarizacion} we show the far field diffraction intensity and polarization structure of the resulting beam for $m=1$, when impinging with a linearly polarized top-hat beam with polarization angles $0^\circ$, $45^\circ$ and $90^\circ$. The result is a beam with variable topological charge ($l=1$, $l=0$ and $l=-1$, respectively). See Visualization 1 for a short movie showing the complete evolution of the output SoP. Intensity distribution remains donut shaped as a Laguerre-Gauss like beam, while polarization structure changes from uniformly vertical to uniformly horizontal passing through hybrid SoPs. In order to plot the polarization ellipses we used a color code based on the form factor $f = b/a$, which is the ratio between the minor $b$ axis and mayor $a$ axis of the polarization ellipse, and whose sign depends on the vector sense of rotation (negative for right-handed and positive for left-handed). In a neighborhood near $f = 0$ we considered polarization to be linear (green), and around $f = \pm1$ we considered polarization to be circular (blue), in any other case the polarization is elliptical (red).

In order to test the applicability of these OAM-tuned beams we used an optical tweezers Matlab toolbox designed by T. Nieminen \textit{et al.} \cite{nieminen07}. This software allows the calculation of optical forces and torques, and can be used for both spherical and non-spherical particles with Gaussian and other beams. We simulated the resulting optical force over a dielectric sphere with radius $0.25\lambda$ (where $\lambda$ is the wavelength) and refractive index $n_p=1.45$ (silica), suspended in a medium with $n_m=1.333$ (water), exerted by the superposition of a vertically polarized Laguerre-Gaussian ($\textnormal{LG}_p^l$) beam with radial index $p=0$ and azimuthal index $l=1$ and a horizontally polarized $\textnormal{LG}_p^l$ beam with radial index $p=0$ and azimuthal index $l=-1$, both focused by means of an objective with a numerical aperture $NA=0.5$. The parameters chosen are typical of optical tweezers experiments, see for example \cite{gecevicius14}.

\begin{figure}[H]
    \centering
    \includegraphics[width=0.5\columnwidth]{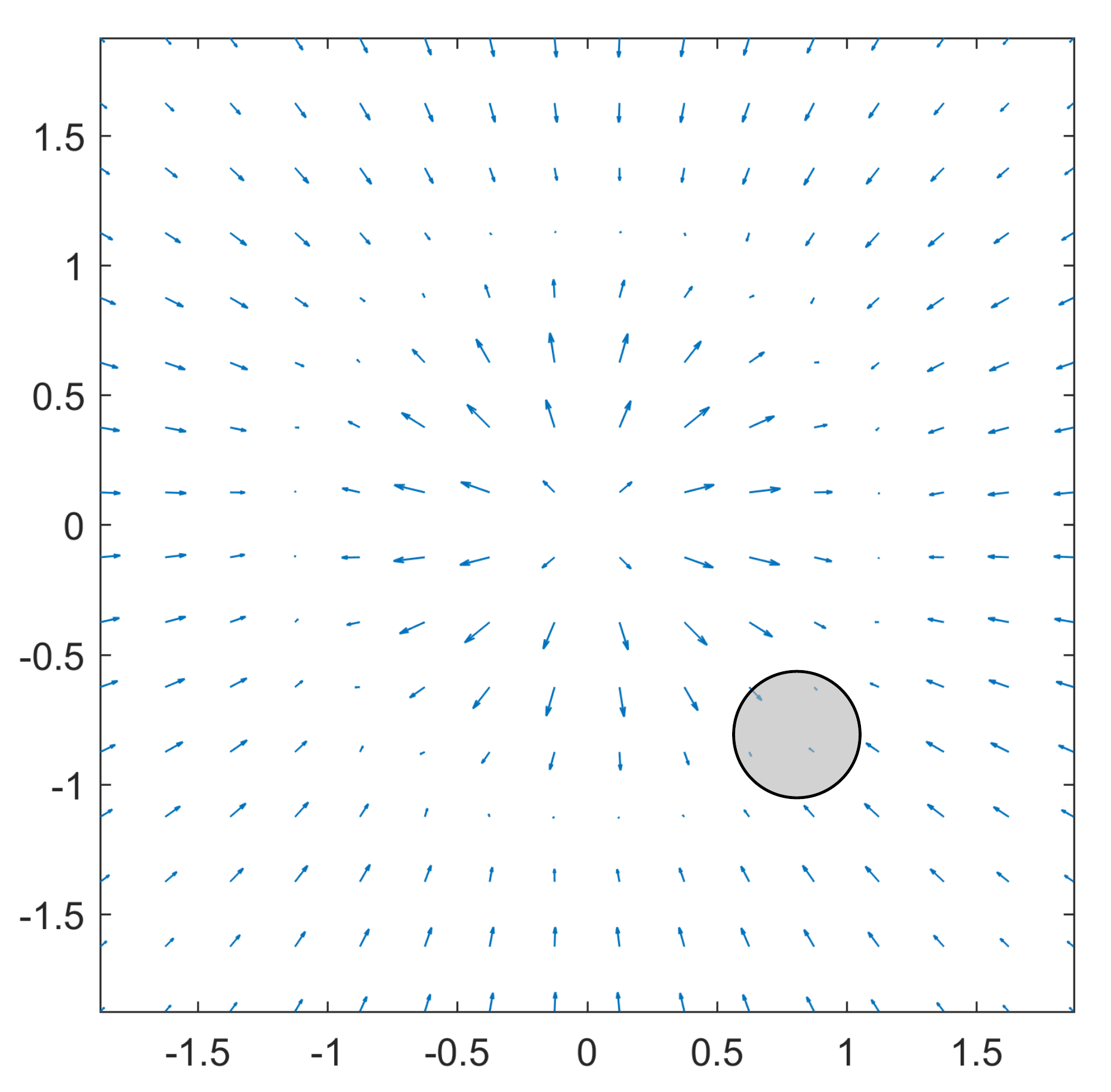}
    \caption{Optical force exerted by a vertically polarized $\textnormal{LG}_0^1$ beam focused with a numerical aperture $NA=0.5$ over a spherical dielectric particle of radius $0.25\lambda$ and refractive index $n_p=1.45$ (silica) suspended in a medium with refractive index $n_m = 1.333$ (water). The particle is at a point of stable equilibrium in the propagation axis (optical trap). Axes are in units of $\lambda$. In Visualization 2 we show the motion of the particle inside the OAM tunable trap, for five different angles of LP.}
    \label{fig:oforce}
\end{figure}

In Fig. \ref{fig:oforce} we show the optical force field in the $xy$ plane obtained when input light is horizontally polarized, i.e. the force exerted over the particle by a vertically polarized $\textnormal{LG}_0^1$ beam. The force is in arbitrary units and axes in units of the wavelength $\lambda$. Optical force in the propagation direction $z$ shows a stable equilibrium point (trap), which does not coincide with the focus for $\textnormal{LG}_p^l$ beams \cite{nieminen07}. The $z$-position of the particle is fixed to this stable equilibrium point in the simulation. The force due to intensity gradient traps the particle in the maximum intensity ring, while the force due to phase gradient gives the rotation around beam axis, as expected \cite{roichman08}.

The OAM tuning of the resulting beam is evident when simulating the motion of a particle due to these optical forces, while varying the incident polarization angle. A short movie showing the evolution of a particle inside the trap with $m=1$ for five different input polarization angles is provided in Visualization 2. This simulation was done by numerically integrating the second order non-linear vector equation of motion
\begin{eqnarray}
   \mathbf{\ddot{x}} = \mathbf{F}(\mathbf{x}) - \gamma\mathbf{\dot{x}},
\end{eqnarray}
where $\mathbf{x} = (x,y)$ is the position vector, $\mathbf{F}$ is the bi-dimensional optical force field calculated with the toolbox, $\gamma$ is a dissipative parameter due to the medium and mass is taken as 1. This equation can be rewritten as a first order system
\begin{eqnarray}
\eqalign{
    \mathbf{\dot{x}} &= \mathbf{v} \\
    \mathbf{\dot{v}} &= \mathbf{F}(\mathbf{x}) - \gamma\mathbf{v}
}
\end{eqnarray}
and solved using an iterative method, like Runge-Kutta \cite{strogatz15}.

\begin{figure}[H]
    \centering
    \includegraphics[width=0.5\columnwidth]{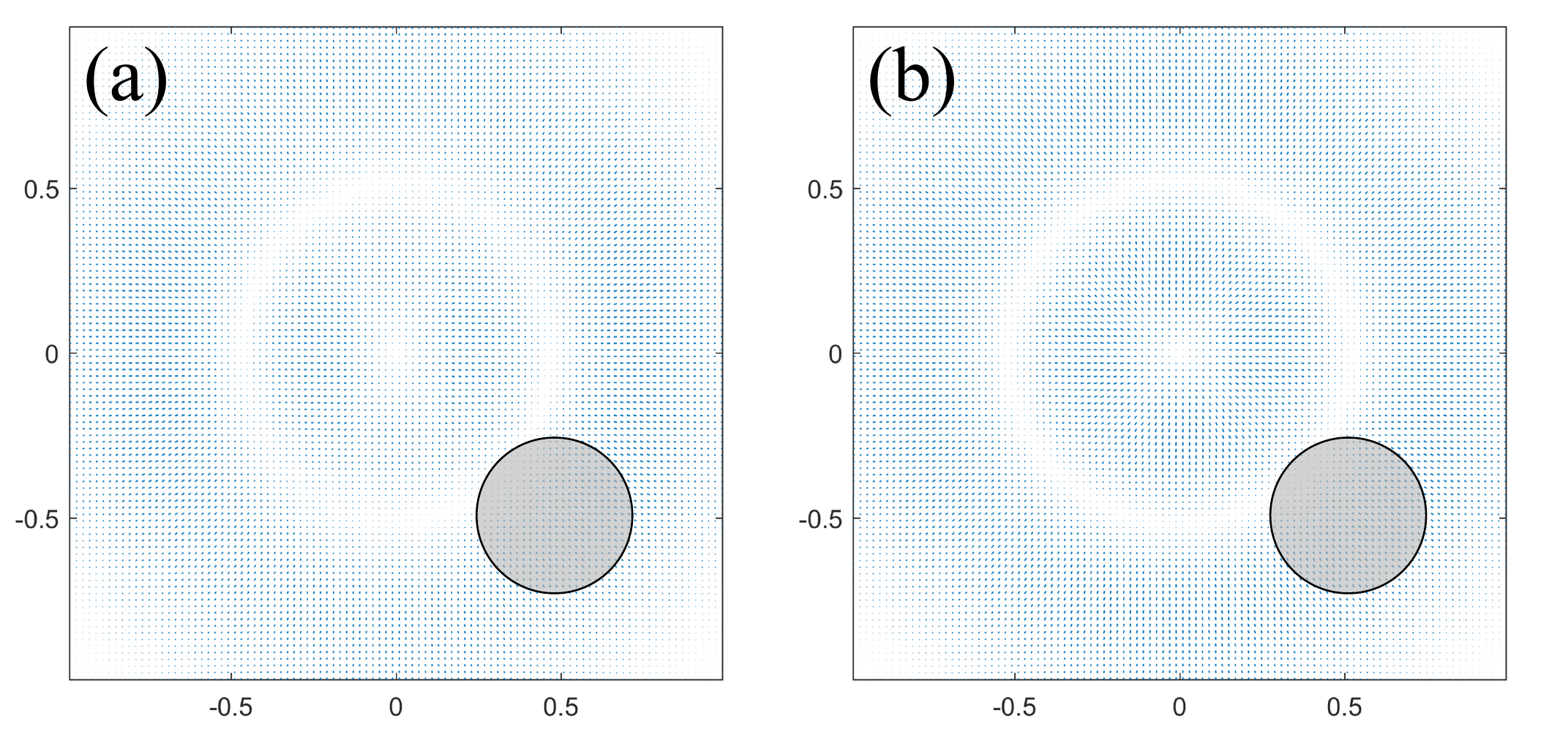}
    \caption{Optical force fields exerted by (a) a vertically polarized $\textnormal{LG}_0^1$ beam and (b) a right circularly polarized $\textnormal{LG}_0^1$ beam, both focused with a numerical aperture $NA=0.95$, over a spherical dielectric particle of radius $0.25\lambda$ and refractive index $n_p=1.45$ (silica) suspended in a medium with refractive index $n_m = 1.333$ (water). The particle is at a point of stable equilibrium in the propagation axis (optical trap). Axes are in units of $\lambda$. In Visualizations 3 and 5 we show the motion of the particle in each case, respectively.}
    \label{fig:oforce_casos}
\end{figure}

This scheme offers a simple way of continuous OAM tuning and shows good results for low numerical apertures. Although, it is known that when focusing a beam with uniform linear polarization the symmetry of the intensity distribution is lost as numerical aperture is increased \cite{dorn03}, due to the unbalanced contribution of the outer regions of the beam. This symmetry breaking is therefore present in the respective optical force field, which causes the loss of the trapping properties and the described behaviours. In Visualization 3 we show the performance of the device when input light is horizontally polarized and output light is focused with a numerical aperture $NA=0.95$ over the particle previously described. Since output beam's polarization is uniformly vertical, upper and lower regions loss intensity at focus, and the optical force field is deformed, as seen in Fig. \ref{fig:oforce_casos}a. As a result, the particle moves slowly through these regions. This can be easily overcame using elliptically polarized light instead.

A beam with uniform circular polarization preserves symmetry at focus, independently of the numerical aperture of the system. So the solution is to adapt the scheme for using orthogonal states of circular polarization instead of linear polarization. That is turning the left circular component of the input beam into right circular, with a phase $\exp(im\theta)$, and turning the right circular component of the input beam into left circular, with a phase $\exp(-im\theta)$. This is exactly what a $q$-plate does \cite{marrucci06}, and is known as the SAM to OAM conversion (STOC) phenomenon.

\begin{figure}[H]
    \centering
    \includegraphics[width=0.5\columnwidth]{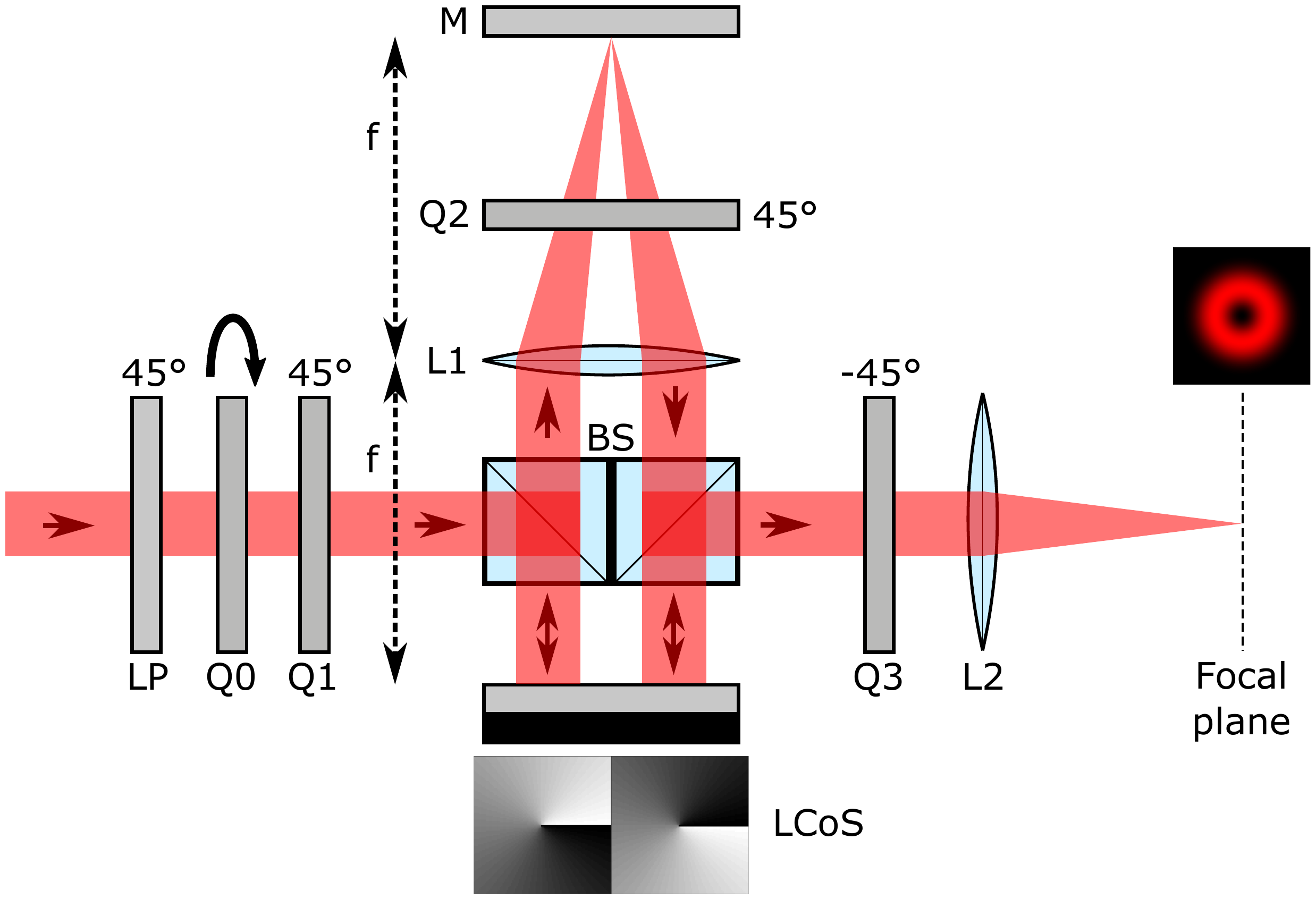}
    \caption{Experimental setup proposed for arbitrary controlled OAM tuning. LP is linear polarizer, Qs are quarter wave plates, Ls are convergent lenses, M is a mirror and BS is a pair of beam-splitters. Orthogonal circular polarization components of the input beam acquire helical phase profiles of opposite handedness, which are programmed onto a phase only PA-LCoS.}
    \label{fig:dispositivo2}
\end{figure}

This result can be managed by introducing quarter-wave plates Q1 (oriented at $45^\circ$) and Q3 (oriented at $-45^\circ$) as shown in the device of Fig. \ref{fig:dispositivo2}. These wave-plates transform the input and output beams accordingly in order to emulate the behaviour of a $q$-plate \cite{vergara19}, adding the desired phase modulation to circularly polarized orthogonal components of the input field. Jones matrix of the device can be written now as
\begin{eqnarray}
\eqalign{
\hat{J}(\theta) &= Q(-45^\circ)\cdot J(\theta)\cdot Q(45^\circ)\\
  &= \left( \begin{array}{cc}
     -\cos(m\theta) & -\sin(m\theta) \\
     -\sin(m\theta) & \cos(m\theta) \end{array} \right)\\
  &= \exp(i\pi)J_{q}(\theta).
}
\end{eqnarray}
This matrix representation coincides with that of a $q$-plate ($J_{q}$) with $q = m/2$, up to a global phase factor \cite{moreno16}.

\begin{figure}[H]
    \centering
    \includegraphics[width=0.5\columnwidth]{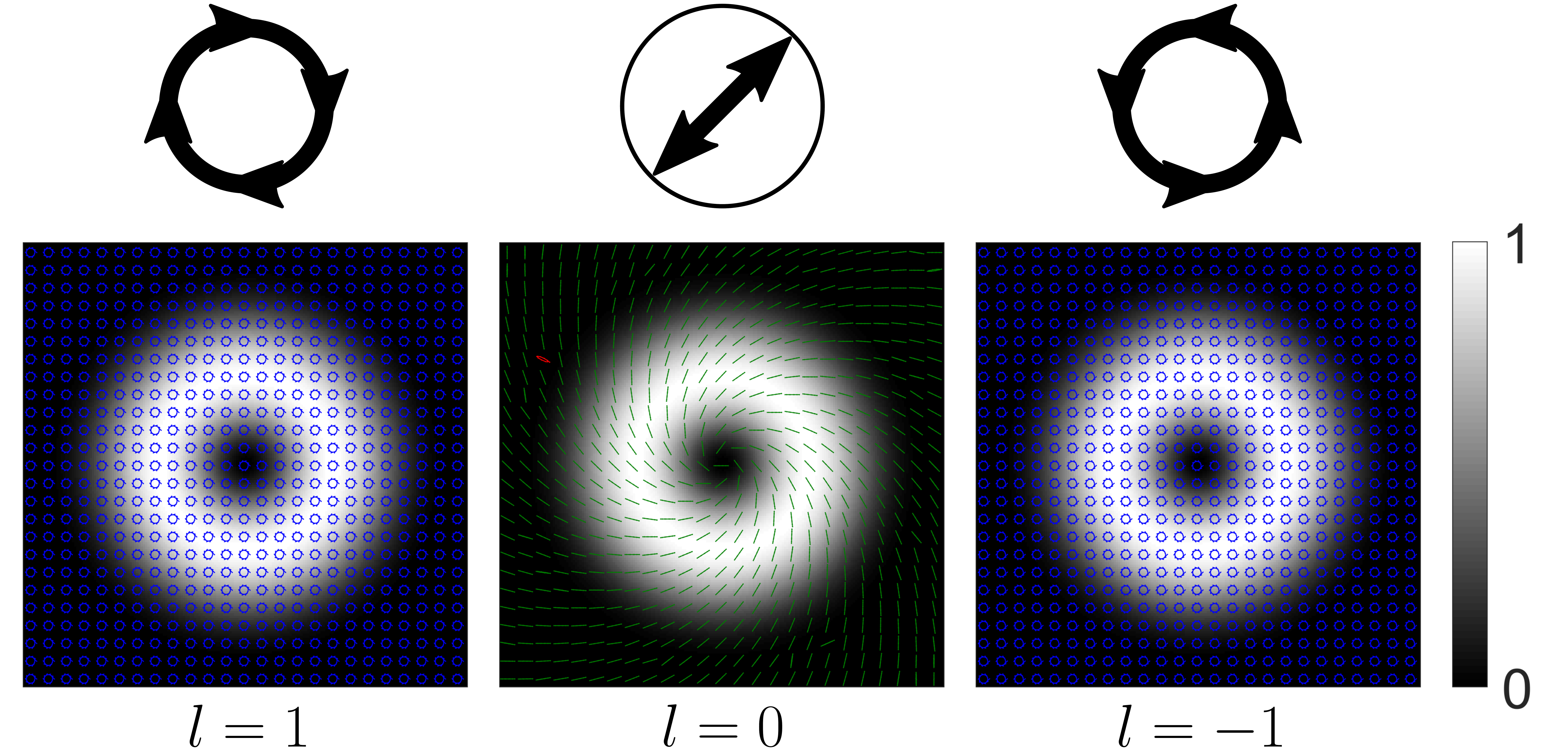}
    \caption{OAM tuning with circular polarization basis. Top: SoP of the incident light. Bottom: Intensity and polarization structure of the OAM tuned beam. Intensity distribution remains, while polarization changes its structure. In Visualization 4 we show the complete evolution of the output SoP.}
    \label{fig:polarizacionC}
\end{figure}

With this scheme the OAM tuning is achieved by continuously changing the ratio between intensity of left and right circularly polarized components of the input field, i.e. its polarization ellipticity. This can be done by a configuration consisting of a linear polarizer (LP) oriented at $45^\circ$ followed by a quarter wave plate (Q0) with variable angle. When Q0 is oriented at $0^\circ$ input light results left circularly polarized, while for a $90^\circ$ orientation it results right circularly polarized, intermediate angles tune the ellipticity of the input SoP, including linear polarization when Q0 is oriented at $45^\circ$. In Fig. \ref{fig:polarizacionC} we show the focused output beam for these three extreme cases: left circular polarized input ($l=1$), linearly polarized input ($l=0$) and right circular polarized input ($l=-1$), all intermediate $l$ values are achieved with elliptic polarized inputs, see Visualization 4 for a movie showing the full transition of the output SoP for this scheme. In Visualization 5 we show a simulation about the motion of a dielectric sphere affected by the resulting optical force field for left circularly polarized input light ($l=1$), focusing with a numerical aperture $NA=0.95$. Now the force field shows cylindrical symmetry, as can be seen in Fig. \ref{fig:oforce_casos}b, and the particle follows a uniform circular motion, in contrast with the linear polarization case discussed earlier.

Summarizing, we have proposed two simple schemes for OAM tuning using a commercially available PA-LCoS, in which the resulting OAM is determined by the input SoP, and shown its applicability by simulating the motion of a dielectric spherical particle due to the OAM tuned beams generated, modeled as superposition of two Laguerre-Gaussian modes with orthogonal polarization states and opposite handedness phase gradients. The principal advantages are the simplicity of the construction and the flexibility given by the LCoS, which allows to change easily the addressed phase, without requiring the fabrication of different elements for reaching higher values of OAM.

\ack{
This work was supported by UBACyT Grant No. 20020170100564BA, and ANPCYT Grant No. PICT 2014/2432. M.V. holds a CONICET Fellowship.}

\section*{References}
\providecommand{\newblock}{}

\end{document}